\documentclass[12pt,a4paper]{article}
\usepackage[dvips]{graphics,epsfig}

\newcommand{\ybox}[2]	{
 \begin{center}
 \resizebox{!}{#1\textheight}
{\includegraphics{#2.eps}}
 \end{center}		}

\begin{document}
\thispagestyle{empty}

\begin{center}
{\large \bf Sensitivity of the Auger Observatory to ultra high energy 
photon composition through inclined showers}
\end{center}
\begin{center}
{\bf  M. Ave$^1$, J. A.~Hinton$^{1,2}$, R. A.~V\'azquez$^3$, \\
A. A.~Watson$^1$, and  E.~Zas$^3$ }\\
$^1$ {\it Department of Physics and Astronomy\\
 University of Leeds, Leeds LS2 9JT ,UK \\}
$^2$ {\it Enrico Fermi Institute, University of Chicago, \\ 
5640 Ellis av., Chicago IL 60637, U.S.\\}
$^3$ {\it Departamento de F\'\i sica de Part\'\i culas,\\
Universidad de Santiago, 15706 Santiago de Compostela, Spain\\}
\end{center}

\begin{abstract}
We report a calculation of the expected rate of inclined air showers induced
by ultra high energy cosmic rays to be obtained by the Auger Southern
Observatory assuming different mass compositions. We describe some features
that can be used to distinguish photons at energies as high as 10$^{20}$ eV.
The discrimination of photons at such energies will help to test some models
of the origin of ultra-high-energy cosmic rays.
\end{abstract}

\section{Introduction}

Uncovering the origin, composition and energy spectrum of the highest energy
cosmic rays is one of the biggest challenges in astroparticle physics.  The
Pierre Auger project is the next step in the search for answers to intriguing
questions about the origin of these particles \cite{augerreport}.  The Auger
Southern Observatory will consist of 1600 water Cerenkov detector stations
(each 10 m$^2$ x 1.2 m deep) on a hexagonal grid of 1500 m spacing overlooked
by four detectors capable of detecting the fluorescence light emitted by the
nitrogen molecules excited by the shower. The array covers a ground area of
approximately 3000 km$^2$ at a mean altitude of 875 g cm$^{-2}$ 
($\sim 1400$~m), near
Malarg\"ue in Mendoza State, Argentina (lat = -35.2$^\circ$, long =
-69.2$^\circ$).  The Auger Southern
Observatory will be able to measure the energy of the incoming cosmic ray
using fluorescence detectors (FD) and thus calibrate the energy inferred from
the surface detectors.
  
The proposal to use the Auger observatories to search for very inclined
showers induced by ultrahigh energy neutrinos \cite{zas} has led to an
investigation of the characteristics of high energy showers at large zenith
angles, i.e. showers arriving at zenith angles larger than
60$^0$~\cite{hplp}. These showers would not be very different from vertical
showers except for the fact that they develop in the upper part of the
atmosphere. As a result the electromagnetic part of the shower, produced mainly
from $\pi^0$ decay, is mostly absorbed well before the shower front reaches
ground level. The muon front propagates through the atmosphere and it gets
attenuated more slowly through pair production, bremsstrahlung, and hadronic
interactions that reduce the muon energy and increase their probability to
decay in flight. Therefore the muon energy spectrum at ground will have a
{\sl low energy} cutoff which increases as the zenith angle rises. 
 As a result the average muon energy at ground level also increases. 
Although the bulk of the overall increase in the average muon energy 
with zenith is due to the rise of the cutoff, there is a smaller 
contribution due to the rise of the pion critical energy (the energy at 
which the pions are more likely to decay than to interact).
These energetic
muons travel making a small angle to the incoming cosmic ray direction but
their trajectories are deflected by the Earth's magnetic field. As a result
the muon density patterns at ground are different from typical densities
measured in vertical showers, reflecting the structure of the shower core and
the geomagnetic field which acts as a kind of ``natural magnetic
spectrometer''.

This work outlines the sensitivity of the Auger Observatory to inclined air
showers induced by ordinary cosmic rays, that is cosmic rays produced by
protons, heavy nuclei, and photons. They constitute the background to neutrino
detection but their observation also provides a significant increase in the
aperture of the array and may improve mass composition studies~\cite{PRL}.
Below, we predict the expected rate for the Auger Surface Array for zenith
angles above $60^{\circ}$ assuming different primary mass compositions:
proton, iron and photons.  Primary photons can interact with the geomagnetic
field before reaching the top of the atmosphere \cite{magnetic}. 
This will have a major impact
on the expected rate of detection for those primaries, as we will show. 
We can take
advantage of this effect to outline a way to identify photon primaries at
energies as high as 10$^{20}$ eV.

The article is organized as follows: in section 2 we describe the procedure to
calculate the expected rate for the Auger array above 60$^\circ$ for primary
protons, giving the expected values for the energy resolution, core error
reconstruction, and multiplicity of stations as a function of the energy: a
detailed explanation of the procedure can also be found in \cite{hplp}; in
section 3 we describe the effect of the geomagnetic interactions in inclined
air showers initiated by photon primaries; in section 4 we present the
expected rate for three primary mass compositions (protons, iron and photons). 
Finally in section 5 we end up with some conclusions.

\section{A simulation of inclined events and their analysis}

The effect of the magnetic field on inclined showers makes the density
patterns at ground level dependent on zenith and azimuth angles, in addition
to their dependence on assumptions that must be made about energy, 
mass composition and
interaction model as is usual in the the case for vertical showers.  The work
presented here relies on the model of the muon density patterns produced by
inclined showers under the influence of the Earth's magnetic field described
in \cite{model}. This model approximates the magnetic distortions of
circularly symmetric muon density patterns generated by simulation in the
absence of magnetic effects. Moreover the model has revealed relatively simple
energy scaling functions for the density patterns once the zenith and
azimuthal angles are determined \cite{rate}.
The model provides a simple and continuous average muon density function which
simplifies fitting procedures and allows us to simulate the expected
sensitivity of the Auger Surface Array for showers with zenith angle above
60$^\circ$.  The magnetic field at Malarg\"ue has been taken to have 
an intensity $B=26~\mu$T, an inclination of $\beta_i=-33.18^\circ$ (pointing 
in the upward direction) and a declination $\beta_d=6.88^\circ$ (eastward). 
A northern site with a larger intensity and higher inclination would have 
a different distortion pattern and results are expected to be slightly 
different particularly in relation to azimuthal angle distributions.

We have simulated 
a subarray containing 91 detectors arranged in five concentric hexagons 
and showers are assumed to fall in the central region to save computing time. 
Results obtained in this subarray are then scaled up to the total array 
surface of 3000~km$^2$. This approximate approach has two important 
implications. The subarray ``radius'' ranges between $r_{sub}\sim 6.5$ and 
7.5~km. This implies that in the transverse plane the 
maximum distance to shower axis is $r_{sub} \cos \theta$ which can be rather 
small for high zenith angles (for example for a zenith of 75$^\circ$ the 
subarray stops sampling at about $1.8$~km). As a result the number of tanks 
that have a signal in the total array will be severely underestimated 
for high zenith and high energy showers. 
Another implication is that no edge effects are taken into 
consideration but this is not expected to affect much neither the shower 
rate nor most of our results because a large fraction of the showers will 
fall well within the array.

With the help of AIRES \cite{aires} (version 2.2.1) and this model, we have 
generated 
libraries with average muon density maps for inclined showers between 
60$^\circ$ and 88$^\circ$ in steps of 2$^\circ$, and for all 
azimuth angles in steps of 5$^\circ$, at a constant energy of 
10$^{19}$ eV. Muon density profiles 
for different primary energies are obtained using the appropriate scaling 
factor which depends on primary composition and hadronic model considered 
\cite{rate}. In this work we consider the QGSJET98 \cite{qgsjet} as the high 
energy hadronic interaction model. To account for the detector response to 
muons of 
different energy and impact angle we have used the WTANK \cite{Wtank} 
program based on the GEANT CERN package \cite{GEANT}. As was previously 
shown in \cite{rate} for Haverah Park detectors, the signal of very inclined 
muons in water \v Cerenkov tanks can be greatly enhanced by direct light and 
muon interactions, which are included in these simulations. 
Between 60$^\circ$ and 70$^\circ$ the electromagnetic component, from $\pi^0$
decay, was directly calculated using Monte Carlo simulation with AIRES and
treated as a small correction to the tank signals, which is ignored 
for $\theta > 70^\circ$.

Using these muon density libraries, and assuming an isotropic cosmic ray 
distribution consisting of protons, and a given parameterization 
of the measured spectrum, we have simulated artificial events in 
the energy range 10$^{18.5}$-10$^{21}$ eV and in the zenith angle range
60$^\circ$-89$^\circ$. We have assumed a simple trigger condition demanding 
5 stations with a density greater than 0.4 vertical equivalent muons (VEM) 
m$^{-2}$. 
For each artificial event the densities at each detector are fluctuated 
according to the Poisson fluctuations in the number of muons, and also 
according to the distributions of light fluctuations in the tank. 
These zenith angle dependent distributions include the response to 
the geometry of the muon tracks, the direct light contribution and 
the catastrophic muon energy losses, namely hard bremsstrahlung, 
pair production or hadronic interactions which become more important as 
the average muon energy rises, as obtained in detailed simulations 
using WTANK. 

Following the steps of \cite{hplp,PRL} the shower parameters of these 
artificial events ($\theta$, $\phi$, impact point position, and energy) are
reconstructed as if they were real data. The procedure is based on a two-step 
fitting process using a maximum likelihood method which is iterated.
 Silent stations are taken into account in the fitting procedure. 
The arrival direction angles are first fitted to the 
signal times in the ``triggered'' detectors and then the energy and 
impact point are varied to make the average muon density map fit 
the actual ``measured'' densities in the tank. Curvature corrections 
to the arrival times are implemented once there is an estimate of the 
core location:  the process is iterated three times for convergence.  

\begin{figure}
\ybox{0.4}{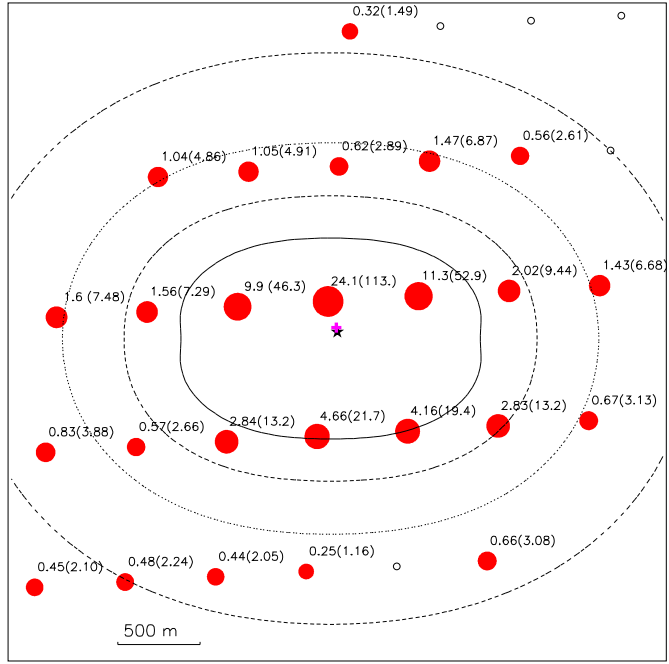}
\caption{Density map of a proton simulated event of 10$^{19.6}$ eV at a zenith
angle of 77$^\circ$ in the plane perpendicular to the shower axis. Recorded
muon densities are shown as circles with radii proportional to the logarithm
of the density. The positions of the best-fit and simulated impact points are
indicated by a star and a cross respectively.  Densities in VEM~m$^{-2}$ are
marked and, in brackets, the actual number of inclined muons that produced
each signal. The $y$-axis is aligned with the component of the magnetic field
perpendicular to the shower axis. Contour levels for 2, 5, 10 and 20 muons per
station are shown for the fit. Detectors not triggered are indicated by an
empty circle.  The original energy was $\log_{10}(E/eV)= 19.52$ and the
reconstructed $\log_{10}(E/eV)=19.6 \pm 0.06$. The original and reconstructed zenith
angles are $\theta = 77^{\circ}$ and $76.5^{\circ}\pm 0.3$ respectively.}
\label{event}
\end{figure}

An example of a reconstructed event in the plane perpendicular to the shower 
direction is shown in Fig. \ref{event}, together with the contours of 
densities that best fit the data. In the figure the array is rotated in the
shower plane such that the $y$-axis is aligned with the component of the
magnetic field perpendicular to the shower axis. The asymmetry in the density
pattern due to the geomagnetic field is apparent.
As anticipated \cite{zas}, in spite of a relatively moderate energy, 
common events can be rather spectacular having over twenty tanks triggered. 

We have estimated the detection rate as a function of primary energy
integrating over all zenith and azimuth angles, using the parameterization of
the energy spectrum of \cite{watson}. This parameterization lies between the
energy spectra recently reported by the AGASA group and the HiRes group
\cite{Sakaki},\cite{HiRes}. Assuming these parameterizations of the energy
spectrum our predicted rate will increase (decrease) up to 20 \%. For
self-consistency, the energy spectrum to be used in a practical application of
the ideas outlined here should be the one derived from the Auger experiment
using the fluorescence technique, as the spectrum derived by this technique
has the advantage of being independent of assumptions of primary mass
composition.

\begin{figure}
\ybox{0.5}{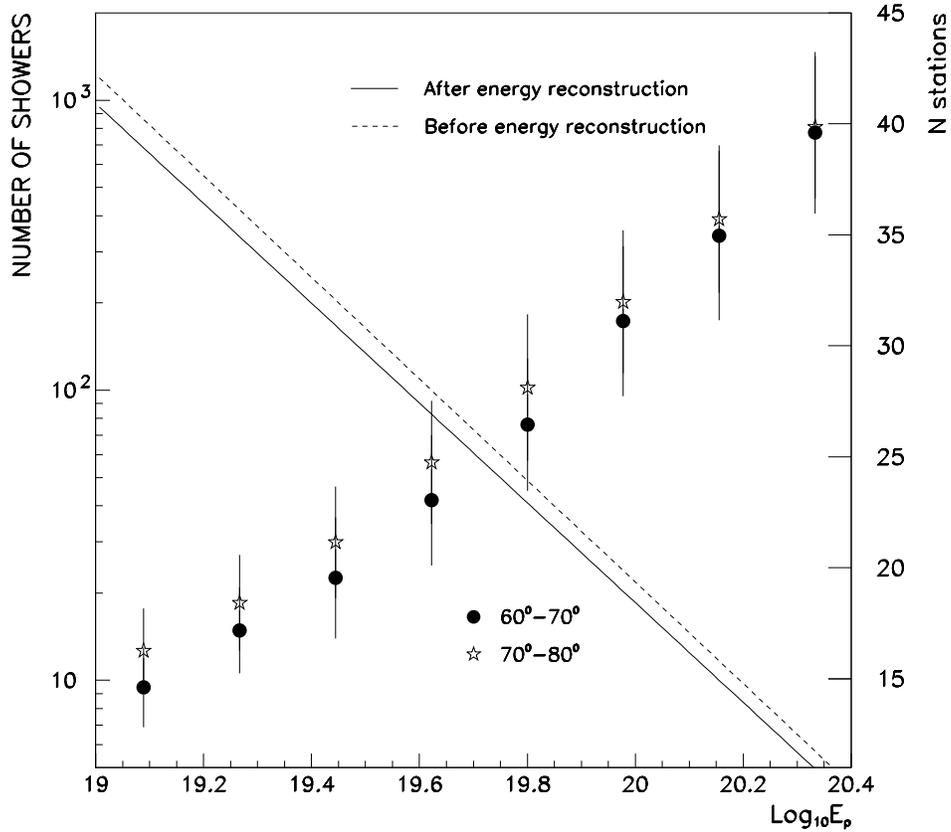}
\caption{Integral spectrum of events per year that trigger the Auger surface
array, before and after energy reconstruction, assuming proton composition and
adopting QGSJET as the high energy interaction model. The average
multiplicity in the number of stations above threshold and its spread 
is also shown for two ranges in the arrival zenith angle 
indicating that the multiplicity is increasing as the zenith rises. 
Hardly any reduction is observed at high energies. This is an artificial 
effect because of the limited number of stations simulated.
Because we are using a reduced subarray, the tank multiplicity given here 
should be considered as a lower bound (see text).}
\label{rate}
\end{figure}

The results are shown in Fig. \ref{rate}.  The dashed line is the integral
rate that could be expected in a 3000~km$^2$ array from the cosmic ray 
spectrum that has been assumed in this work. 
The continuous line represents the reconstructed
spectrum. We have made quality cuts to the 
reconstructed events, namely the downward error in the energy was required
to be less than 60\% and the $\chi^2$ probability for the energy and time 
fits have been forced to be greater than 1\% for the accepted events. 
The difference in normalization, which amounts to 
$\sim$ $20 \%$, is due to the 
combined effect of the quality cuts, systematic effects and the effect of 
the finite energy resolution of the detector.

The number events  expected per year, assuming a pure proton 
composition, is $\sim 1000$ above $10^{19}$ eV and $\sim 18$ above $10^{20}$
eV after making these quality cuts. It should be noted that these numbers are
 obtained using the parameterization of the energy spectrum of \cite{watson},
 where no Greisen-Zatsepin-Kuzmin (GZK) cutoff is indicated.

The multiplicity of detectors with a signal above 0.4~VEM~m$^{-2}$ as a
function of energy for the range of angles simulated is also shown in
Fig. \ref{rate}.  For a given energy the multiplicity is much larger than for
vertical showers.  Two factors contribute to this effect: the reduction of the
effective spacing between stations in the shower plane, and the flatter
density profile of horizontal air showers compared to vertical.

As a shower becomes inclined the earliest and
latest tanks that are above threshold can be quite far apart corresponding to
different depths in shower development. Due to the additional atmosphere 
traversed an attenuation can be expected for the late arriving particles. 
This was studied for muon 
distributions which is the dominant component in inclined showers in 
\cite{model,lorenzo} and for electron distributions in \cite{dova}. 
If the number density of 
muons changes significantly over such a distance it can represent a
significant correction to the densities derived in our work. 
It should be remarked that no attenuation of the shower particles across
the array, has been taken into account in this work. 
As it is shown in \cite{model}, such attenuation effects become 
important at
extreme zenith angles $> 87^\circ$, but for most of the zenith angle interval
discussed in this article the attenuation, which is less important than in 
the vertical case, can be ignored in a first approximation. 
As an example, simulations show that for a 30$^\circ$ 
shower, water Cerenkov density at 1000 m from the shower core in the shower
plane is attenuated by a $\approx$ 30 \% when reaching ground level. At 
70$^0$ this factor is less than 15 \%. 
The main reason for this is that in the
vertical direction the signal is dominated by electromagnetic particles from
$\pi^0$ decay, while in the horizontal direction the signal at ground level is
dominated by high energy muons and secondary electromagnetic particles from
decaying muons.  The scale of the attenuation is roughly determined by the
interaction length, which for photons and electrons is orders of magnitude
smaller than that for muons.

\begin{figure}
\ybox{0.3}{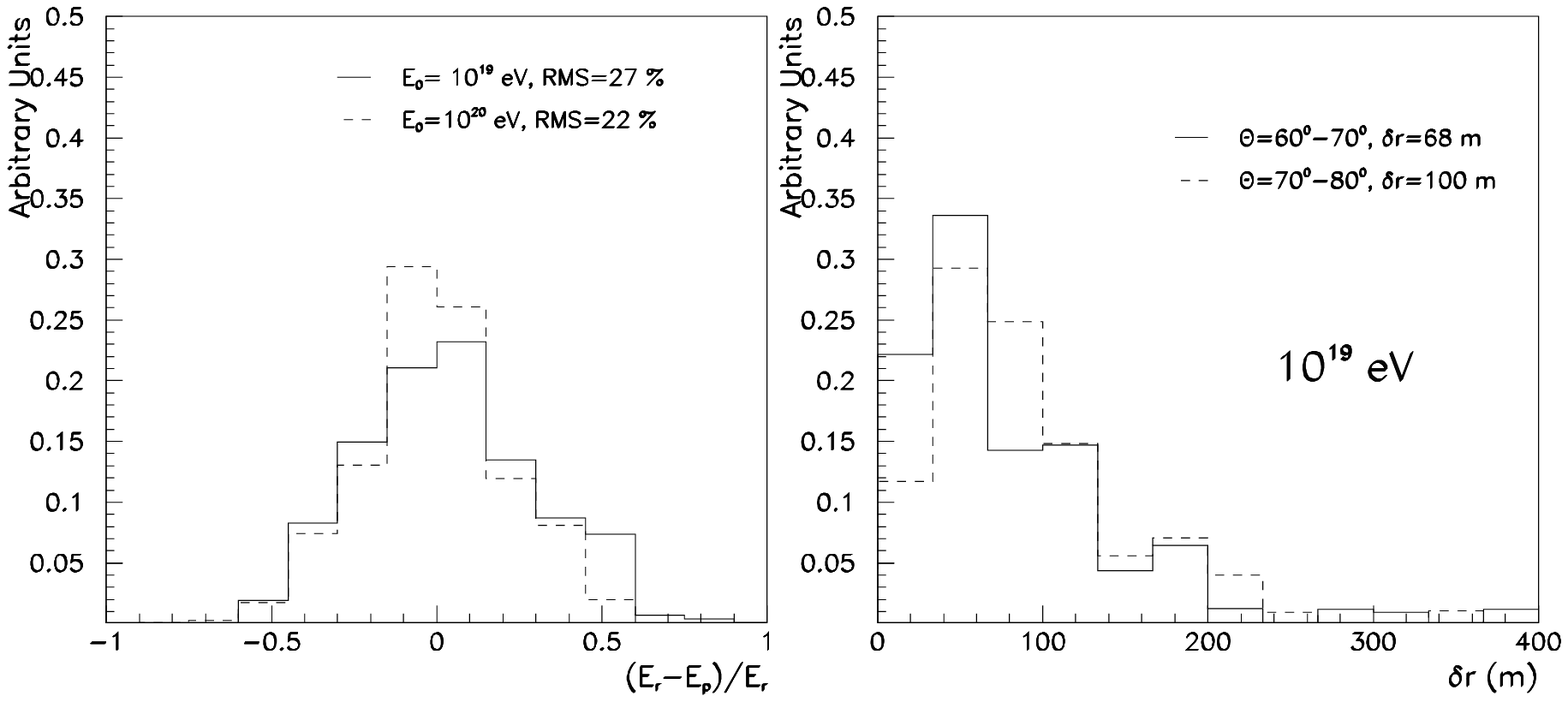}
\caption{Left panel: energy resolution integrated for all zenith in two
energy bins. A flat energy distribution is assumed for each graph. Right
panel: distribution of the difference between the real and the reconstructed
impact point positions in the energy range $10^{19.5}-10^{20}$ eV for two 
zenith angle ranges.}
\label{energycomp}
\end{figure}

We should also point out that in spite of using average density maps, 
statistical fluctuations of the shower densities are considered in this work. 
These are complicated effects that include fluctuations in the ratios of 
energy transported by charged and neutral pions in the hadronic interactions 
as well as fluctuations in depth of maximum. 
As obtained in simulations it typically represents a $20\%$ effect 
on the overall muon number density normalization for proton primaries and 
it is rather independent of primary energy \cite{hplp}. 
Much of the fluctuation corresponds to the first interaction.
In the left panel of Fig. \ref{energycomp} we show the energy resolution for
two energy ranges. 
Two factors contribute to the energy resolution calculated
in this work: the variation of the number of muons at ground due to shower to
shower fluctuations ($\approx 20 \%$ for all energies) and the muon size
reconstruction error (which evolves from 20 \% at 10$^{19}$ eV to 10 \% at
10$^{20}$ eV).  Hence, at 10$^{20}$ eV the energy resolution is dominated by
shower to shower fluctuations.  
The errors in the reconstruction of the impact points of the analysed events 
are also shown in Fig. \ref{energycomp}. As an example it is $\sim 70$ m at 
60$^\circ$ and it has a small variation with zenith angle due to 
the different effective spacing of the detectors when projected on to the 
transverse shower plane. 

\section{Inclined showers induced by primary photons}

The recent analysis of inclined data obtained at Haverah Park has opened a new
window to establishing bounds on photon composition at ultrahigh energies
\cite{PRL}. As photon showers have far fewer muons than hadron showers, the
expected rate of inclined showers created by photons detected with a surface
array is reduced relative to the rate from protons or nuclei primaries.  This
is because the detection of horizontal showers induced by photons and hadrons
by surface detectors is practically only sensitive to the muon component. The
determination of the spectrum of cosmic rays through vertical shower
measurements using the fluorescence technique, which is fairly independent of
the primary composition together with the inclined shower rate thus allow 
measurements of composition \cite{hplp,PRL}.

Following a similar procedure to that described in the previous section we
obtain in this section the predicted integral spectrum for photon primaries
for the Auger Observatories. There are a number of potential problems that
could compromise the capabilities of the Auger Observatories to establish
bounds or measure photon composition through inclined event rates, namely the
Landau-Pomeran\v cuck-Migdal~\cite{LPM} effect, the rise in the
photoproduction cross section~\cite{halzen} and the interactions of photons
with the geomagnetic field of the Earth~\cite{magnetic,stanev,bertou}.  We
discuss these effects in detail to show to what extent it can be expected that
the Auger detector will be sensitive to photon primaries.

\subsection{General features of inclined photon showers}

\begin{figure}
\ybox{0.4} {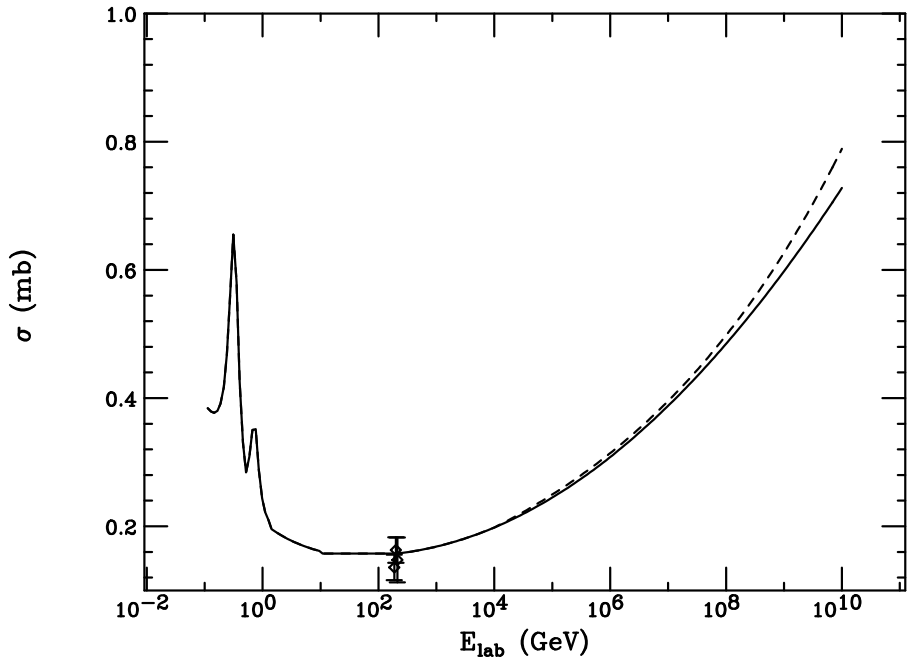}
\caption{Photon--Proton photoproduction cross section as a function of
the photon energy. Continuous line represents the parameterization of
the AIRES code. The dotted line is the Block {\it et al.}
parameterization \cite{halzen}. Both curves are normalized at low
energy. Points show data from HERA \cite{H1,ZEUS}.}
\label{photocross}
\end{figure}

The dominant process for muon production in an electromagnetic cascade is 
photoproduction:
\begin{equation}
\gamma + nucleus \rightarrow hadrons \\
\label{photoprod}
\end{equation}
When photoproduction occurs the reaction products are essentially like those
of a pion-nucleus interaction. Muons originate from the decay of the produced
pions and kaons and their progeny in the resulting hadronic subshowers. The
cross section for photoproduction has been measured up to $\sim 10^4~$ GeV
\cite{H1,ZEUS} for the incident photons in the laboratory frame. Above the
resonance region (see Fig. \ref{photocross}) the cross section is about 100
$\mu$barns per nucleon, and rises slowly for photon energies above $\sim 10~$
GeV. The corresponding cross section on an air nucleus is $\sim$
1.1 mb, obtained by scaling $\sigma_{\gamma A}\sim A^\alpha 
\sigma_{\gamma p}$ where $\alpha\sim 0.9$. The ratio
of the pair production to the photohadronic cross section gives the
probability of a hadronic interaction to occur. This ratio is at 10 GeV:
\begin{equation}
R=\frac{\sigma_{\gamma \rightarrow hadrons}}
{\sigma_{\gamma \rightarrow e+e-}} \sim 2.8~10^{-3} \\ 
\label{ratio}
\end{equation}
The ratio grows with the incident photon energy, because of the rise in the 
photoproduction cross section, and it is expected to be
$\sim 10^{-2}$ at $10^{19}~$ eV. Fig.~\ref{photocross} shows the measured
photoproduction cross sections at different energies, together with the 
standard parameterization 
used by the AIRES 
code \cite{aires} and a recent parameterization given by Block {\it et al.} 
\cite{halzen} from a combined analysis of $pp$, $p \gamma$ and $\gamma \gamma$ 
interactions. 
As pointed out in ref.\cite{halzen}, the photoproduction cross section 
is close to the saturation limit at the HERA energies and cannot grow with
energy faster than the currently used parameterizations. Therefore, 
the Block {\it et al.} parameterization may be considered as an upper limit
to the photoproduction cross section. 
Other potential sources of muons are smaller than this. Muon pair
production, in which a pair of electrons is replaced by a pair
of muons is suppressed by a factor  $m_e^2/m_{\mu}^2 \sim 2~10^{-5}$ 
\cite{PLBgaisser91}. 
Hadron production by electrons also contributes less than
photoproduction because the process has to occur through exchange of
a virtual photon, and the energies of the produced hadrons are small.

The probability of photoproduction versus pair production 
is further enhanced at high energies when the Landau-Pomeranchuk-Migdal 
(LPM) effect takes place \cite{LPM}. The 
LPM effect is a collective effect of the electric potential of several atoms 
and it tends to suppress the pair-production and bremsstrahlung cross 
sections for energies above a given value $E_{LPM}$, which is inversely 
proportional to the medium density. 
It starts to become relevant for the development of photon induced air 
showers at primary energies above $10^{19}$ eV for vertical 
showers \cite{sciutto}, a value somewhat above $E_{LPM}$. This energy rises 
for large zenith angles corresponding to shower development higher up in 
the atmosphere where the air is thinner. 
We have found from simulations using AIRES code that for 
the zenith angle range being discussed, the LPM effect has a mild impact
on the total number of muons produced by photon showers ($<$ 25 \% at
a zenith angle of 60$^\circ$). 

It has been shown before that to a very good approximation the number of 
muons produced by a proton and iron primary rises with energy 
according to a simple scaling law \cite{hplp}: 
\begin{equation}
N_\mu=N_0~(E/10^{19})^{\beta}, \\
\label{Escaling}
\end{equation}
where $N_0$ (the normalization, here taken at $10^{19}$~eV) and $\beta$ (a
constant parameter of order 0.9) are fixed numbers for a given zenith angle,
hadronic interaction model and mass composition. The approximation is 
also valid for
photon primaries although a number of differences are worth a brief
discussion.

\begin{figure}
\ybox{0.4} {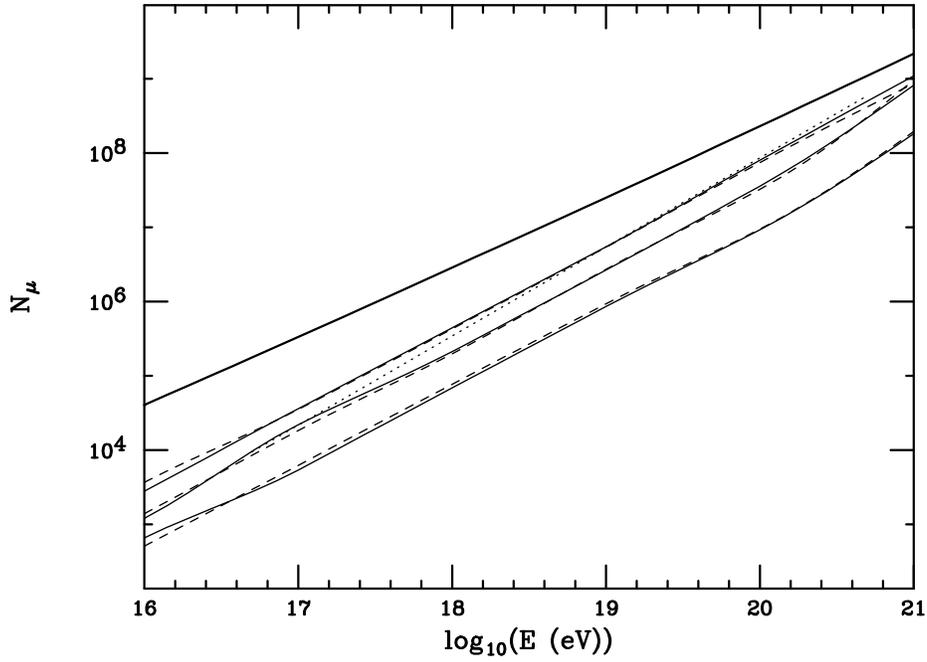}
\caption{Number of muons produced in inclined showers as a function of 
primary energy for proton showers (upper continuous line) and photon showers 
for $60^{\circ}$,$70^{\circ}$,$80^{\circ}$, from top to bottom, using the 
standard AIRES parameterization (continuous) and the Block {\it et al.} 
parameterization (dashed). Also shown for $\theta=60^{\circ}$ is 
the $\beta=1.2$ parameterization used in our previous work (dotted line).}
\label{scaling_ph}
\end{figure}

We have simulated batches of 100 photon induced air showers using AIRES code
at different zenith angles and energies, in the absence of magnetic field,
using the two different extrapolations of the photoproduction cross sections
which are shown in Fig.~\ref{photocross}, namely the standard parameterization
of AIRES and the parameterization of ref \cite{halzen}.  In
Fig. \ref{scaling_ph} we plot the average number of muons generated by a
photon as a function of energy for different zenith angles as obtained in the
simulation. We note that the absolute number of muons of a photon shower is
less than that of a hadron shower as expected from the ratio of probabilities
in Eq.~\ref{ratio}. However the number of muons in a photon-induced shower
rises with energy faster than in a proton shower so that differences between
the two become smaller as the energy rises. This is also qualitatively
expected from the behaviour of the photoproduction cross section with
energy. We also show for comparison an approximate scaling law using
$\beta=1.2$ which we have used before as a conservative estimate
\cite{hplp,PRL}.

We can observe in Fig.~\ref{scaling_ph} a steepening kink of the curve at 
the highest energies. This is due to the LPM effect which makes 
the showers develop deeper and further enhances the ratio $R$. 
The rise in $R$ is a potential problem for the method. If the average number 
of muons in a photon shower becomes sufficiently close to that of a proton 
shower, no conclusion could be extracted regarding photon composition since 
a photon shower that photoproduces in the first interaction is practically 
undistinguishable from a proton shower of the same energy. 

On the other hand, the difference in the number of muons produced for the two
parameterizations used for the photoproduction cross section is small which
gives an idea of the uncertainty involved in this calculation (see
Fig.~\ref{scaling_ph}).  This is not surprising since it is clear that the
number of muons in a photon shower is dominated by processes at lower energies
closer to the region where data constrain the cross section.

Fluctuations in the number of muons in photon-induced showers are rather
different from ordinary cosmic ray showers. If the first interaction of the
incident photon happens to be hadronic (probability R$\sim$ 0.01 at $10^{19}$
eV) then the shower is indistinguishable from a hadronic shower.  We can thus
expect a distribution for the total number of muons in a photon shower with
a long tail in the region of showers with large number of muons, close to
those of a proton shower. These non-gaussian tails can affect the detection
rates and for that reason we have implemented fluctuations in the number of
muons in photon shower using the distributions obtained in the simulations.

\subsection{Interactions with the geomagnetic field}

It was pointed out in \cite{magnetic,stanev,bertou} that photons above 
$10^{19}$ eV have a large probability to convert into an electron-positron 
pair in the 
presence of the Earth magnetic field before entering the atmosphere. The 
strength of this interaction depends on $EB_{\perp}$ where $E$ is the photon 
energy and $B_{\perp}$ is the transverse magnetic field. Therefore the 
probability of interaction will depend on the direction of the incoming 
photon with respect to the Earth frame. For a given zenith angle it is a 
maximum when the photon trajectory intercepts the Earth's magnetic axis, 
which roughly corresponds to the geomagnetic south at the Malarg\"ue site. 
This is because at a given distance to the center of the Earth the dipole 
field is strongest along the magnetic axis. For the same reason it is clear 
that the field will be largest for higher zenith angles. 
This azimuthal asymmetry was also pointed out by Stanev and Vankov
\cite{stanev} and by Bertou {\it et al.}  \cite{bertou} to be a potential 
way to establish a photon contribution in the highest energy cosmic rays.

The point we want to make is that this interaction reduces the potential 
problems that could arise because of the increase in the photoproduction cross 
section and therefore ``safeguards'' the method of using the inclined shower 
rate for composition studies also for the highest energies. 
If such an interaction is deemed to happen an electron-positron pair is 
created 
that will radiate strongly by synchrotron radiation. A large number of 
photons is thus produced and some of them may give secondary pairs provided 
they still have sufficient energy. Instead of a single particle 
entering the upper atmosphere we end up with a spectrum of gammas and 
one or a few electron positron pairs adding up to the primary photon energy. 

\begin{figure}
\ybox{0.4}{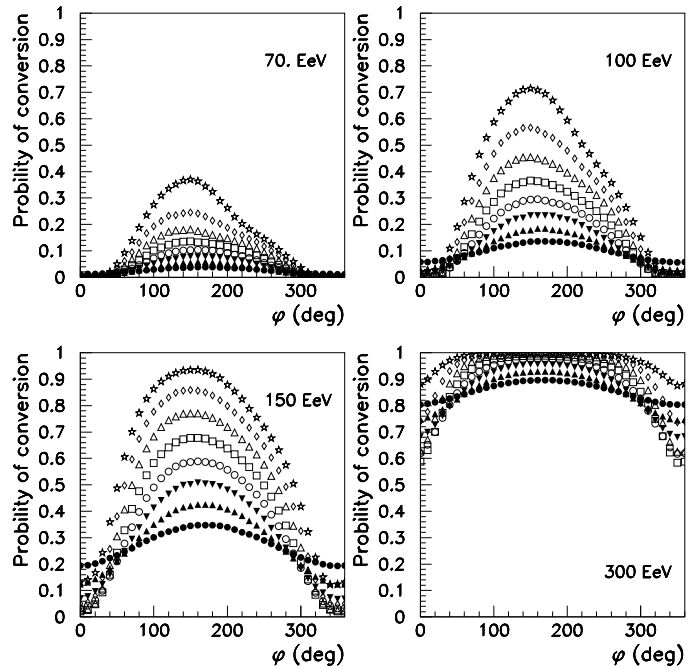}
\caption{Probability of photon conversion in the geomagnetic field as a
function of azimuth angle for four different energies and eight different 
zenith. In each graph the eight curves are displayed corresponding, from 
bottom to top, to zenith angles from 10$^\circ$ to 80$^\circ$ in steps of 
10$^\circ$. An azimuth angle of 0$^\circ$ corresponds to a photon 
arriving from the geomagnetic north. 
}
\label{prob}
\end{figure}

To study this effect we have devised a Monte Carlo to simulate the cascading
of photons in the geomagnetic field. The photon is injected at 20000 km from
the top of the atmosphere along the incoming direction and tracked in steps of
100 km. The values of the interaction length for pair production are taken
from \cite{erber}.  Fig. \ref{prob} shows the probability of conversion for
photons at different energies, zenith and azimuth angles. When a photon is
converted, the resulting electron and positron are tracked in steps of 2 km,
and the number and energy of the photons emitted via synchrotron radiation
is calculated using the spectral distribution in \cite{erber}. The photons
produced by magnetic bremsstrahlung are tracked and the probability of
conversion calculated in each step.  In this way, we obtain the spectrum of
electrons, positrons and gammas at the top of the atmosphere. 

The relevant parameter for the inclined shower rate is the total number of 
muons in the shower. When the photon converts in the magnetic field the 
number of muons at ground will be given by a sum over the contribution of the 
corresponding particle distribution: 
\begin{equation}
N_\mu=\sum_{i} N_0~(E_i/10^{19})^{\beta}, \\
\label{nmuconv}
\end{equation}
where $E_i$ is the energy of particle $i$.
Meanwhile for an unconverted photon it is given by Eq.~\ref{Escaling}. Since 
the slope parameter in this equation, $\beta$, is greater than one
the number of muons is reduced when the photon interacts with the Earth's 
magnetic field relative to when it does not. 

\begin{figure}
\ybox{0.4}{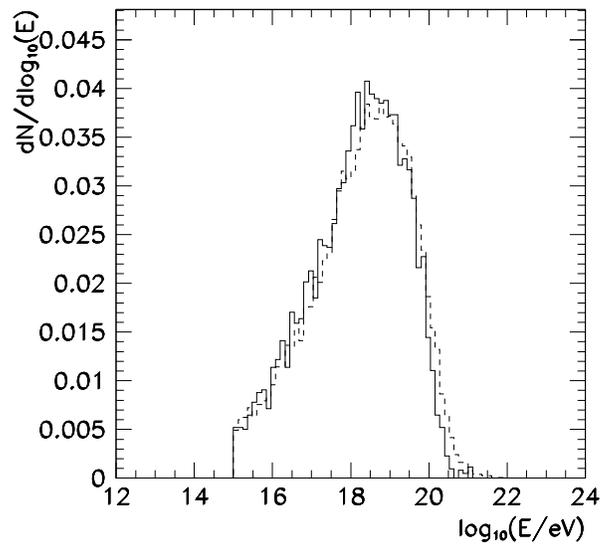}
\caption{Differential energy spectrum of electromagnetic particles resulting
 of the cascading in the geomagnetic field of primary photons at two different
 primary energies $10^{21}$ eV (continuous) and $10^{22}$ eV (dashed).} 
\label{spec_uni}
\end{figure}

It is of some interest to comment on a few facts about this conversion 
procedure. The strength of the Earth's magnetic field is such that efficient 
conversion starts at energies of order $10^{20}~$eV. 
When the photon energy is well above this energy, the photon 
is always converted, giving rise to a sort of 
universal photon spectrum. 
The shape of the resulting spectrum is governed by magnetic bremsstrahlung 
and it is independent of the primary energy and only mildly dependent on the 
arrival direction; only the total number of 
particles changes as can be seen in Fig.~\ref{spec_uni} for two different 
primary energies.  
The spectrum displays a low energy tail and a maximum at around $10^{19}~$eV 
and can be parameterized by: 
\begin{equation}
\frac{dN_{em}}{d\log E}=k~\left[\frac{E}{E_c}\right]^\gamma 
\left[1+\frac{E}{E_c}\right]^\delta,
\label{spec_eq}
\end{equation}
where $\gamma$=0.24, $\delta$=-1.7, $E_c$=10$^{19.76}$ eV, and $k$ is a 
normalization constant which due to energy conservation 
scales linearly with the primary photon energy.
At very high energies there is a sharp cutoff in the spectrum due to the 
fact that all photons of such energies convert in the magnetic field. 
Thus at very high energies, when the photons convert, we can equivalently 
talk about an effective slope parameter of (Eq. \ref{Escaling}) $\beta_{\rm eff}=1$. 

At this point we can use the method described in the previous section to
generate artificial events in the Auger array. We use the conversion 
probability for the photon showers to predict their behavior.  
These events are then
reconstructed as if they were protons, and an equivalent energy ($E_p$) is
calculated. 

It is interesting to note that, as a coincidence, the energy at which 
photon conversion becomes efficient is also that at which the 
LPM effects start to show up in shower development, 
$\approx 10^{19}$ eV \cite{sciutto} for a vertical air shower. 
When a photon is converted a quick look at the spectrum shown in 
Fig.~\ref{spec_uni} shows that most of the particles that will reach 
the top of the atmosphere and initiate the shower will be below the LPM 
threshold. The magnetic field of the Earth shields the atmosphere and the 
probability that showers develop in the atmosphere with strong LPM 
suppression becomes very small. The effective shielding is strongest at 
high zenith angles because the energy at which the LPM effect starts to 
take up is higher.

This is fortunate because, as Fig.~\ref{scaling_ph} shows, if a high energy 
photon reaches the top of the atmosphere, the number of muons produced in the 
corresponding shower could become rather large and close to that of proton 
induced showers. 

\section{Composition analysis with HAS}

\begin{figure}
\ybox{0.5}{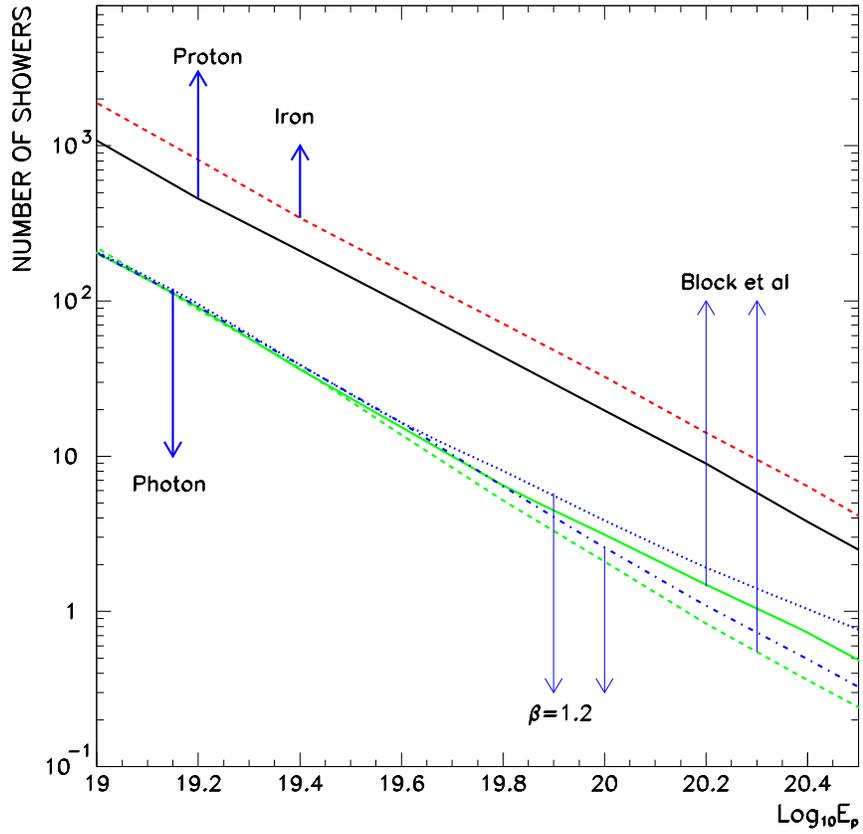}
\caption{Integral $E_{p}$ spectrum of events per year triggering the Auger
Surface Array ($\theta > $ 60$^0$) for three assumptions of primary
composition. The expected spectra for converted and unconverted photons are
also displayed.  The two sets of spectra are calculated using two different
assumptions of the rise with energy of the number of muons produced in photon
showers.}
\label{rate_mass}
\end{figure}

In Fig. \ref{rate_mass} we show the integral energy spectra obtained from the
artificial events under three different assumptions for the primary
composition (protons, iron, and photons) using the cosmic ray parameterization
given in \cite{watson}. In all the cases the artificial events are
generated with muon density maps obtained for each specific primary, and then
reconstructed as if they were protons, obtaining an equivalent energy ($E_p$).
For primary photons we have calculated the spectra switching on (off) the
interaction with the geomagnetic field. 
We show the prediction assuming the Block {\it et al.} photoproduction 
cross section and a parameterization used in \cite{PRL}. This
parameterization is based on AIRES simulations but the authors set the 
 value of $\beta$ to be 1.2 to be conservative.

Three important features are clear in this picture:

\begin{itemize}

\item The expected rate for primary photons above 10$^{20}$ eV is reduced 
30 \% when geomagnetic interactions are taken into account. 
   
\item The two photoproduction cross sections used predict a rate for photon 
primaries $\sim$~10 times smaller compared to the predictions for other 
hadronic primaries. 
  
\item 
Geomagnetic interactions modify the slope of the spectrum for primary 
photons at energies above 10$^{19.6}$ eV. 
\end{itemize}

According to current wisdom and partly because of the geomagnetic effect 
the uncertainty on the photoproduction cross sections at high energies has 
a small effect on the muon number at ground level.  
The predicted distortion of the longitudinal shower development for photon 
primaries because of the LPM effect 
is reduced because of the geomagnetic interactions. 
Assuming that the vertical flux is well measured independently of mass 
composition, as expected to be achieved with a fluorescence detector, 
the rate of inclined showers could be used to impose strong constraints 
on primary photon composition assuming a given hadronic model. 

The hadronic model uncertainty can change the normalization of the 
spectra shown in Fig. \ref{rate_mass}. Therefore this method is not adequate 
to impose constraints on both mass composition and hadronic models at the 
same time. It should be used in combination with vertical 
measurements on mass composition using the same hadronic model. 
However, the differences in muon densities between the different hadronic 
models available are $\approx 20\%$, which is comparable to the difference 
between densities produced by proton and iron primaries, and therefore, 
these primaries cannot be resolved by using the overall rate measurements. 
Nevertheless the difference between photon and hadronic primaries are much 
larger than 20 \% so that this method can, on its own, impose severe 
constraints on the photon content of the highest energy cosmic rays.

Currently the ultra high energy data suggest a mostly hadronic composition 
\cite{PRL,Teshima01},
and the inclined showers rate is well described by proton primaries at energies
above $\sim 10^{19}$ eV. 
If there is a change of composition to a photon dominance at higher energies, 
the rate of inclined showers will be much lower. 
After a few years of operation of the Auger Observatory these features will
help to establish bounds on the flux of photons at energies as high as 
10$^{20}$ eV. 

\section{Conclusions}
\label{conclu}

Inclined showers will be seen in the Auger Observatory as spectacular events
with as many as 30 or 40 hit detectors. 
We have calculated the approximate rate of inclined showers with zenith 
angle exceeding $60^\circ$ expected to be observed at the
Pierre Auger Observatory. This rate increases the aperture of the observatory
by almost a factor 2. Assuming a pure proton composition, there will
be over 1000 well reconstructed events above $10^{19}$ eV, with a mean error energy
$\sim 25$~\%. 
The rate is sensitive to composition. If photons were dominant at high energy,
the rate would be an order of magnitude smaller than if they were 
protons or nuclei allowing for a clear discrimination of the two cases.
Uncertainties in the physics at very high energies have implications for our 
results on the detailed quantitative predictions but these uncertainties 
have little impact on the previous conclusions. 

There are other signatures of the presence of photons (see for instance 
ref.~\cite{bertou}). Surely the combination of the inclined shower 
rate measurements together with vertical flux determinations and detailed 
analysis of the expected photon signatures will be a great step in the 
establishment of the overall photon rate at very high energies with the 
forthcoming data from the Auger experiment.
 
\section*{Acknowledgments} 
We thank F. Halzen, M. Teshima, and K. Shinozaki for discussions and 
P.~Billoir and S.~Sciutto for very useful comments after careful reading 
of the manuscript. 
We also thank F.~Halzen for bringing up the issue of photoproduction 
uncertainty. This work was partly supported by a joint grant from the
British Council and the Spanish Ministry of Education (HB1997-0175), by Xunta
de Galicia (PGIDT00PXI20615PR), by CICYT (AEN99-0589-C02-02 and) by MCYT 
(FPA 2001-3837) and by PPARC(PPA/9/5/1998/00453). The work of R.A.V. is 
supported by the ''Ram\'on y Cajal'' program. We thank the 'Centro de 
Supercomputaci\'on de Galicia'' (CESGA) for computer resources.

\end{document}